# Topological Photonics on a Small Scale

Dmitry V. Zhirihin[1] and Yuri S. Kivshar[1,2]

[1]School of Physics and Engineering, ITMO University, St. Petersburg 197101, Russia
[2]Nonlinear Physics Center, Research School of Physics, Australian National University
Canberra ACT 2601, Australia

Abstract

The study of topological phases of light suggests novel opportunities for creating robust optical structures and on-chip photonic devices which are immune against scattering losses and structural disorder. However, many recent demonstrations of topological effects in optics employ structures with relatively large scales. Here we discuss the physics and realisation of topological photonics on small scales, with the dimensions often smaller or comparable with the wavelength of light. We highlight the recent experimental demonstrations of small-scale topological states based on arrays of resonant nanoparticles and discuss a novel photonic platform employing higher-order topological effects for creating subwavelength highly efficient topologically protected optical cavities. We pay a special attention to the recent progress on topological polaritonic structures and summarize with our vision on the future directions of nanoscale topological photonics and its impact on other fields.

# 1. Introduction

Nanophotonics studies the behavior of light on nanometer scales as well as the interaction of nanometer-sized objects with light. Traditionally, the field of nanophotonics was closely associated with metallic components that can focus and transport light by means of surface plasmon polaritons.[1] A new field of dielectric resonant metaphotonics expands substantially the horizons of the subwavelength optics, and it helps to design efficient photonic devices based on new physical principles.[2] Recently emerged study of topological phases of light provides unique opportunities for creating new photonic systems protected from scattering losses and structural disorder, usually termed as *photonic topological insulators*.[3-11]

General interest to the study of topological effects in optics is driven by a grand challenge of creating robust structures supporting localization and propagation of light. Granted with topological protection, advanced functionalities of novel topological devices are expected to be achieved in active, tunable, and nonreciprocal structures. For implementation of photonic topological insulators at the nanoscale the all-dielectric platform based on resonant dielectric nanoparticles looks very promising. This is because the dielectric nanostructures may support strong electric and magnetic Mie resonances,[2] and they have low material losses which are usually destructive for topology. Based on this platform, we can build topological optical structures with preserved time-reversal symmetry. The preserved time-reversal symmetry is preferential in optics because in optics it is hard to deal with magnetic materials if we want to create devices compatible with integrated optical circuitry and nanophotonics.

In this *Perspective*, we discuss topological effects on a small scale and highlight the recent experimental demonstrations in topological nanophotonics for control of light with subwavelength optical structures. More specifically, here we discuss *three types of* *topological systems* realized experimentally for various photonic platforms on a small scale. All such structures can support highly localized topology-empowered subwavelength optical modes with nontrivial topological properties.

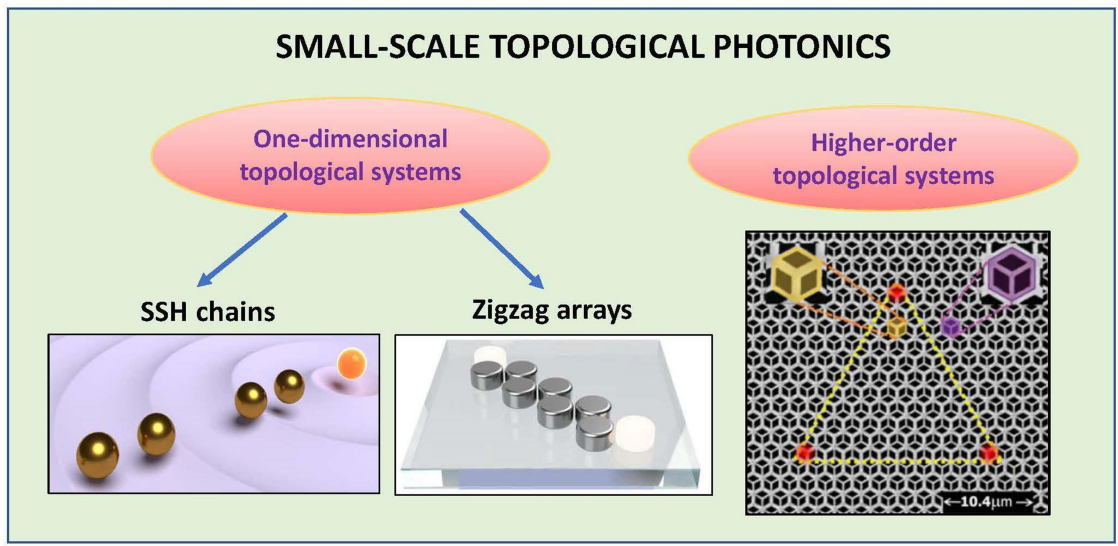

**Figure 1. Three types of photonic structures** that can support subwavelength topological states: dimerized (Reproduced with permission.[15] Copyright 2018, American Chemical Society. Further permissions related to the material excerpted should be directed to the ACS.) and zigzag (Reproduced [78]) arrays of resonant nanoparticles and a dielectric metasurface (Reproduced with permission [46], Copyright 2021, Wiley) with a triangular cavity supporting higher-order topological states (corner states).

*The first system* is an array of resonant nanoparticles described by the one-dimensional Su–Schrieffer–Heeger (SSH) model (see **Figure 1**). This model originates from the pioneering studies of topological charge localization in chains of conjugated polymers,[12,13] and we use this polymer platform to explain the fundamentals of the SSH model. In many realisations of this model (see. e.g., Refs. [14,15] to cite a few), the SSH model describes a kind of "diatomic array" composed of identical elements, as shown schematically in **Figure 1**. The array is characterised by two alternating coupling strengths between adjacent resonators and such binary bonds introduce the hybridization of the individual modes into symmetric/antisymmetric states, creating the characteristic energy splitting. Specific choice of the boundary gives rise to *a topological mode*. At the interface between the two mirror parts, the lattice supports a topological mode, as explained below.

*The second type* of topological systems we discuss here is based on a zigzag array of resonant particles.[16] Some experimental realizations include dipole plasmon resonances in metallic nanodisks[16,17] and quadrupole Mie resonances in spherical dielectric particles.[18] In this case, the polarization-dependent hybridization between the modes supported by the particles gives rise to a pair of degenerate localized states at the edges of the structure. These edge states have been directly visualized in the near field scanning optical microscope,[17] and they can be excited selectively by changing the polarization of the incident plane wave. Emergence of the polarization-degenerate edge states is shown to be a generic feature of the zigzag symmetry, realized for both plasmonic and dielectric nanoparticles, as well as for arrays of coupled optical cavities.

For *the third type* of topological systems, we select the examples of the recently emerged systems classified as *higher-order topological insulators* (HOTIs).[19-23] In contrast to conventional topological insulators, these structures support the states two and more dimensional lower than the original system itself. For example, zero-dimensional (0D) localized corner states arise in two-dimensional HOTI, in addition to one-dimensional (1D) topological edge states, as shown in **Figure 1**. Capabilities of more flexible control over topological states provided by HOTIs attract a lot of attention from different fields of physics, including mechanics,[24] acoustics,[25] electric circuits,[26] and electromagnetics.[27] Importantly, the same analogy could also be extended to 3D systems, but here we focus mainly on two-dimensional structures often associated with the so-called *optical metasurfaces*.

The structure of this *Perspective* is the following. In *Section II* we give a brief description of the three models with nontrivial topological properties. Then in *Section III* we introduce different experimental platforms for implementing those models in small-scale topology. *Section IV* is devoted to discussions of the interplay between nonlinear effects and topology on small scales. In *Section V,* we highlight the recent results on topological exciton-polaritons.

Finally, in *Summary and Outlook* we discuss some future perspectives of small-scale topological photonics.

## 2. Topological Properties

### 2.1. SSH Model

We start with summarizing the theory of the standard SSH chain describing, in the tight-binding approximation, an array of identical resonators with staggered nearest-neighbor couplings *t* and *t'*, as shown in **Figure 2a**. In this model, long-range interactions are neglected. Inequality of coupling strengths *t* and *t'* provides the dimerization of the lattice such as the unit-cell composed of two cites ($a_i$ and $b_i$, where *i* is a number of the unit-cell in the lattice), and it leads to opening of a topological gap. The Hamiltonian describing this system possess chiral symmetry, and is written as:

$$\widehat{H} = \sum_i t c^\dagger_{A,i} c_{B,i} + t' c^\dagger_{A,i+1} c_{B,i} + \text{h.c.},$$

where *t* and *t'* are the coupling coefficients, $c^\dagger_{\frac{A}{B},i}$ ($c_{\frac{A}{B},i}$) is the creation (annihilation) operator on A or B sublattice cite of the *i*-th unit-cell.

It is important to emphasize that there appear two possible choices of the unit cell in the one-dimensional (1D) SSH lattice, as shown in **Figure 2a** by grey boxes. When we consider an infinite lattice, it provides the same band structure for both cases, but the type of truncation becomes crucially important for finite structures. The difference between the unit cells has a topological origin, and it can be characterized by the topological invariant, usually called *winding number W*[28] which takes only integer values and can be linked to the Zak phase[29] divided by π. Non-zero values of the winding number characterize non-trivial topology. The unit cell of strongly coupled resonators (see **Figure 2a**, left panel *t>t'*) exhibits a trivial topological phase, because the winding number vanishes, *W = 0*. On the other hand, the unit cell with 'unbroken' weak coupling (see **Figure 2a**, middle panel *t<t'*) is characterized by *W*

= *1*, manifesting the presence of a nontrivial topological phase. When the termination of a finite structure is chosen to be topologically trivial, no edge states in the energy spectrum appear (see **Figure 2a**, left panel). Topologically non-trivial termination gives rise to the localized topological edge states pinned to zero energy (see **Figure 2**, middle panel). The field profile of the edge state has a maximum at the boundary, and it decays exponentially in the lattice.

From the practical point of view, it is extremely important to consider interfaces between dissimilar systems showing many fascinating physical phenomena. Especially, this is useful in photonics, where the fabrication of artificial periodic structures is relatively simple. The study of domain walls between topologically dissimilar structures demonstrates another way of creating topological edge states. In the case of the SSH model, it could be achieved by connecting two chains with different winding numbers. The interface between the lattices due to the *bulk-boundary correspondence* (the principle stating that the number of topological modes at the interface between two systems is equal to the difference of their topological invariants) supports a localized topological state, as shown in **Figure 2a** (right panel) pinned to zero energy. Such topological interface states are called Jackiw-Rebbi states.[30]

The standard SSH model is one of the major cornerstones of the topological band theory in photonics, and it can be generalized in different ways to describe even richer physics. Considering next-nearest-neighbor coupling terms, the extended SSH model possesses more topological phases than the standard one.[31] Moreover, many interesting phenomena arise with a nonlinear generalization of the SSH model. Thus, introducing Kerr-like nonlinearity leads to a self-induced topological transition for varying intensity, and it modifies topological edge states dramatically, in comparison with their linear counterparts.[32-34] Another important nonlinear example is the SSH model with interacting pair of bosons giving rise to novel types of the doublon (bound photon pairs) edge states.[35, 36]

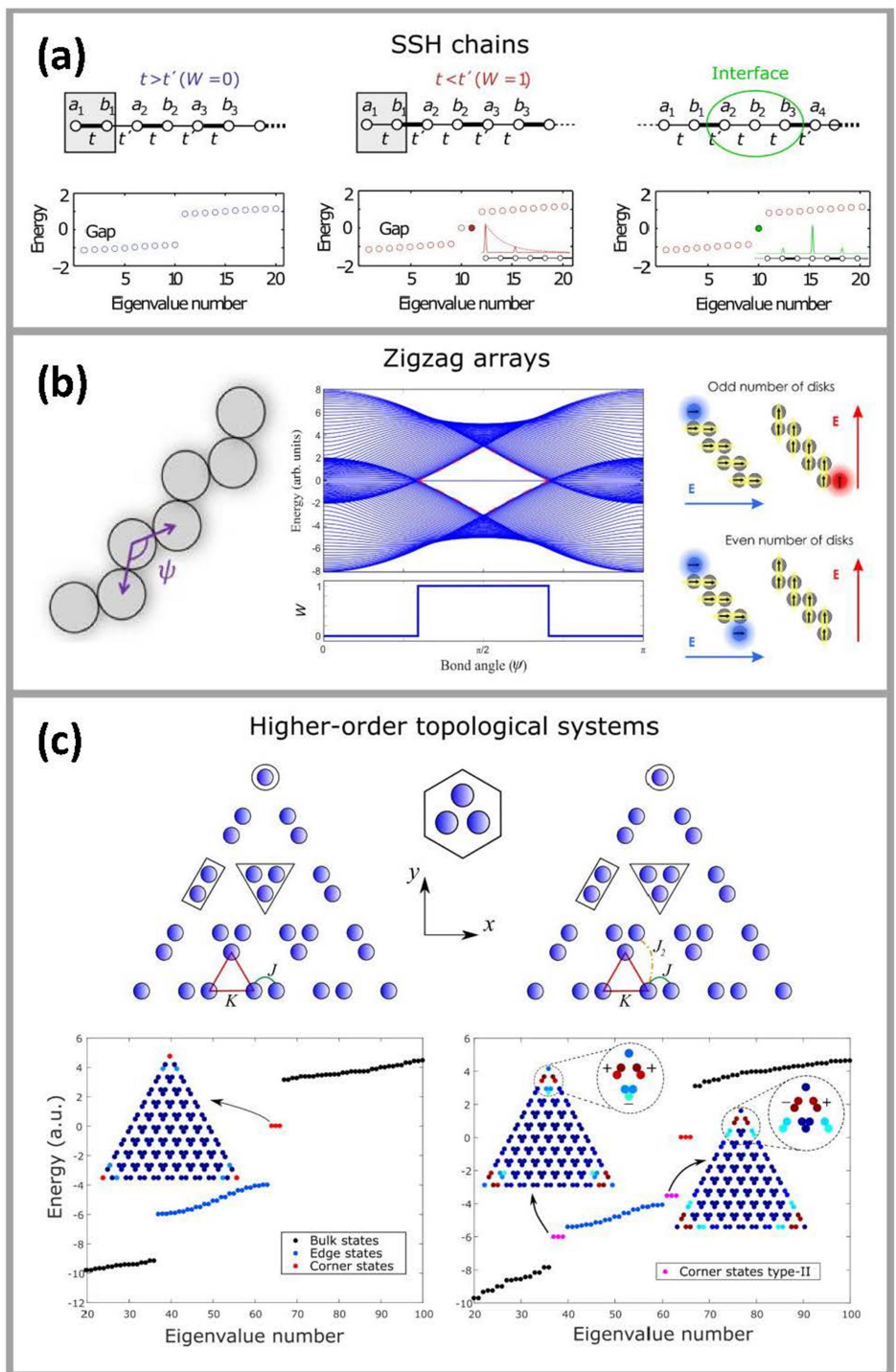

**Figure 2. Topological properties**. Three basic systems supporting subwavelength topological states as described in this Perspective: (a) a dimerized chain of plasmonic or dielectric particles (Adapted by permission from Springer Nature Customer Service Centre GmbH: Springer Nature Photonics[68], Copyright 2017), and (b) a zigzag array composed of resonant dielectric nanoparticles (middle panel - reproduced with permission.[18] Copyright 2015, American Physical Society), (left panel - Reproduced under terms of the CC-BY license.[39] Copyright 2020, Published by the American Physical Society), (right panel - reproduced [78]), (c) a photonic metasurface with the kagome lattice supporting different types of higher-order topological localized corner states (Adapted by permission from Springer Nature Customer Service Centre GmbH: Springer Nature Photonics[42], Copyright 2019).

**2.2. Zigzag Model**

Some years ago, a new type of topological model supporting edge states was suggested theoretically[16] and realized experimentally[17,18] in zigzag arrays of resonant nanoparticles, a sketch of such an array is presented in **Figure 2b** (left column). Despite the similarity with the SSH model, the physics of the zigzag arrays is wealthier and more general. To describe theoretically the topological properties of the zigzag arrays, we employ the coupled-mode approach originating from the existence of resonant modes excited in each nanoparticle. The chiral Hamiltonian for the zigzag model can be written in the form:

$$\widehat{H} = \sum_{j,v} \omega_0 a_{jv}^\dagger a_{jv} + \sum_{(j,j'),v,v'} a_{jv}^\dagger V_{vv'}^{(j,j')} a_{j'v'},$$

where $\omega_0$ is the resonance frequency, $a_{jv}^\dagger$ ($a_{jv}$) is the creation (annihilation) operator for the multipolar eigenmodes with the polarization $v$ at the j-th element, $V$ is a coupling matrix for the co-polarized and cross-polarized modes of the zigzag elements which depends on the bond angle $\psi$ between the adjacent zigzag elements, as described in more details in Ref.[18]. For simplicity, we consider only in-plane excitation by the normal incident electromagnetic waves and only the nearest-neighbor interactions of the zigzag elements in the array. The typical calculated spectrum of the above Hamiltonian for a finite chain is shown in **Figure 2b** as a function of $\psi$. The bulk spectrum becomes gapped within the interval $|\psi - \pi/2| < \sin^{-1}|(t_{||} + t_\perp)/(t_{||} - t_\perp)|$, where $t_{||}$ and $t_\perp$ are the co- and cross-polarized coupling coefficients, and it is characterized by the emergence of zero-energy edge modes. A non-zero value of the winding number, which can be calculated for the zigzag structure,[18] proves the existence of a non-trivial topological phase within the gapped region (see **Figure 2b**, middle column).

Important distinction of the zigzag system from the SSH model is the emergence of topological edge mode for both odd and even numbers of elements (**Figure 2b**, right column) owing to the degeneracy of its linearly polarized eigenmodes. The mutual orientation of neighbor multipole moments (e.g., dipole moments shown in **Figure 2b**, right column) provides either weak or strong interaction between particles, and weakly coupled edge particles support topological edge modes for each of the cases. Thus, for the array with even number of elements, topological edge modes emerge at both edges for a proper excitation, and they are co-polarized. On the other hand, for odd number of elements, the edge modes are cross-polarized which allows to excite selectively edge modes by manipulation of the light polarization that also gives rise to the photonic spin-Hall effect.[37] It is important to mention that the topological edge modes can also be classified by the types of multipolar Mie resonances enabling broader opportunities to control over the light at the nanoscale.[38]

Another fascinating topological phenomenon is revealed from analysing of a generalized zigzag model with staggered couplings which might be obtained by varying the distances between the elements.[39] When the staggered potential is introduced into the zigzag model, the topological phase diagram acquires an additional topological phase with the winding number $W = 2$ characterized by four-fold degenerate topological edge states. Further investigations of such systems with disordered bond angle $\psi$ manifest disorder-induced topological phase transitions for two cases: between the phases with the winding numbers $W = 0$ and $W = 1$, and between the phases with the winding numbers $W = 1$ and $W = 2$.[39]

### 2.3 Higher-Order Topological Insulators

The field of HOTIs originates from the pioneering works[19, 20] where the authors formulated a theoretical framework for higher-order electric multipole moments of crystals and demonstrated that systems with quantized quadrupole/octupole moments host topologically protected 0D corner states. Such systems could be described by tight-binding models with introduced negative intra- and inter-cell couplings. The first demonstration of a photonic quadrupole topological system supporting localized corner states was realized with an integrated silicon photonic platform based on ring resonator waveguides,[40] but the relative size of the unit-cell was much larger than the operating wavelength.

The challenge was overcome by realizing higher-order photonic topological structures based on kagome lattices [41, 42] and later, on honeycomb lattices.[43] Earlier theoretical studies revealed[44] the possibility to obtain higher-order topological states in the $C_n$-symmetric systems with only positive couplings. The theoretical approach is the same for all such systems, and below we just briefly describe the case of the $C_3$-symmetric kagome lattice. Unperturbed kagome lattice (when intra- and inter-cell couplings are equal, $K = J$) has the Dirac-like gapless bulk spectrum. Making coupling coefficients unequal opens a band gap.[45] It turns out that for a finite lattice composed of expanded unit-cells ($K < J$), topological edge and higher-order topological corner modes emerge within the band gap at the boundaries of the system, as shown in **Figure 2c** (left column). For the opposite case ($K > J$), there is no boundary modes in the system.

More importantly, photonic higher-order topological systems possess richer physics than their condensed matter counterpart due to the inherent long-range interactions. Introducing the next-nearest-neighbor coupling $J_2$ into the kagome lattice model, we can find new types of localized corner states with symmetric (lower) and antisymmetric (upper) field profiles branching from the edge-state continuum, as shown in **Figure 2c** right column. Such type-II corner states were initially obtained theoretically, and then observed experimentally for microwaves,[42] and later – in optics [46] by employing near-field scanning microscopy. This discovery unveils the diversity of 0D photonic topological states, and it opens wide opportunities to employ them in nanophotonics.

## 3. Different Platforms for Topological Nanophotonics

Recent advances in the theoretical description of many fundamental phenomena in physics were achieved by applying topology tools. Initially, topological features were found in condensed matter physics which defined further focus of their investigations. However, the unique functionalities of the systems with topological phases supporting robust edge transport gave rise to active research in different directions, including mechanics[47-50] and acoustics.[51-54] Recently, general concepts of topology-empowered photonics[9, 11] demonstrated new non-conventional and even counterintuitive concepts of light manipulation.

We have introduced above three theoretical models describing the basic topological physics underpinning the existence of localized states in the SSH and zigzag arrays, as well as HOTIs in metasurface structures. Here, we discuss different implementations of those models in realistic photonic platforms with a special focus on the small-scale design and the existence of subwavelength topological edge and corner states.

To the best of our knowledge, the first experimental observation of 0D topological states in traditional subwavelength plasmonic structures was reported for a zigzag chain of metallic nanodiscs.[17] By using near-field scanning optical microscopy, a selective polarization-dependent excitation of topological edge modes was demonstrated. The theoretical study of dimerized plasmonic nanoparticle arrays suggests the emergence of topological edge plasmon modes at the interface between diatomic chains.[55] Long-range interactions with retardation and radiative losses may provide advanced functionalities beyond the traditional SSH model, but experimental demonstrations are still missing. Similarly, the schemes to realize higher-order topological photonic nanostructures based on plasmonic systems have been proposed only theoretically.[56, 57]

The realization of small-scale topology can also be based on a semiconductor platform. Semiconductors have a large refractive index in the near-infrared regime that allows to realize systems with a wide photonic band gap. One of the possible designs of semiconductor-based photonic topological structures is an array of coupled microring resonators (see **Figure 3a**) in an InGaAsP multilayer quantum well structure.[58, 59] In such structures staggered coupling strength is controlled by variation of distances between adjacent rings, mimicking the SSH model and enabling a nontrivial topological phase with topological edge modes.

Another important types of semiconductor-based topological structures are photonic crystal nanocavity arrays (nanobeam).[60, 61] An example of such a nanobeam is depicted in **Figure 3b**, and it is composed of an SSH-patterned GaAs slab. The unit-cell contains two airholes which size difference gives rise to topological bandgap opening. 0D topological edge mode emerges at the interface between two topologically distinct photonic crystal nanobeams. Furthermore, photonic crystal nanocavity arrays allow to study two-dimensional topological structures with 0D corner states,[62] including higher-order topological states.[63]

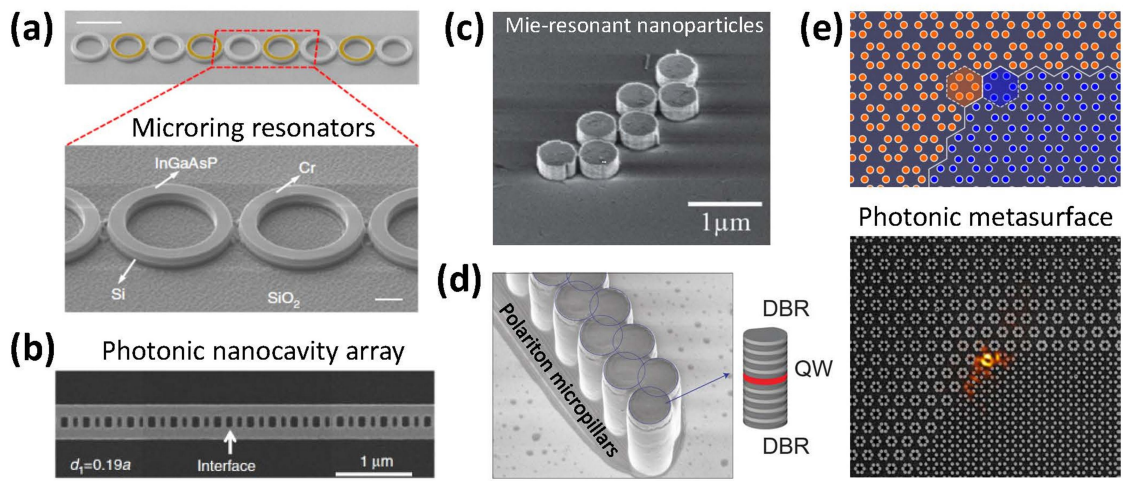

**Figure 3. Different platforms for topological nanophotonics**: (a) SSH model realized with microring resonators (Reproduced under terms of the CC-BY license.[58] Copyright 2018, Published by Springer Nature), (b) an array of photonic-crystal nanocavities (Adapted under terms of the CC-BY license.[60] Copyright 2018, Published by Springer Nature), (c) a zigzag array composed of dielectric nanodisks (Reproduced with permission.[38] Copyright 2017, Wiley), (d) an array of exciton-polariton micropillars (Adapted by permission from Springer Nature Customer Service Centre GmbH: Springer Nature Photonics[68], Copyright 2017), (e) dielectric metasurface supporting topological edge and corner states (Reproduced with permission[43], Copyright 2021, American Chemical Society).

Recently emerged field of dielectric nanophotonics provides an excellent platform for realizing subwavelength topological modes. In contrast to plasmonic particles, dielectric particles have low Ohmic loss and low-heating properties, and they can possess multiple Mie-resonances of electric and magnetic types. Such properties make dielectric-based structures as a remarkable alternative due to advanced functionalities over light control especially for nonlinear effects.[64] Topological phase transition accompanied with the emergence of both electric and magnetic photonic topological edge states were observed in zigzag chains of dielectric nanoparticles,[38] as shown in **Figure 3c**. Furthermore, the dielectric platform allowed to investigate advanced physics in disordered zigzag nanodiscs with staggered-coupling and disorder.[39] Dielectric nanostructures can also be good candidates for studying higher-order topological phenomena.

Thus, silicon-on-insulator topological metasurfaces of the kagome-type lattice were experimentally realized and investigated at optical frequencies.[46] Direct observation of higher-order topological states, including long-range interaction induced type-II corner states, were performed using scattering scanning near-field optical microscopy. Furthermore, metal-dielectric hybrid structures represented as Mie-resonant honeycomb metasurface (**Figure 3e**) can host 0D higher-order topological states at optical frequencies. [43, 64]

Recently, exciton-polariton systems (**Figure 3d**) gather significant attention as a promising platform for topological photonics. [65-73] Due to the hybrid light-matter nature, such systems show fundamental differences from the other topological photonic systems. Below, we describe topological exciton-polariton systems in more details.

## 4. Emission and Lasing from Topological States

The field of topological photonics emerged initially from the replication of topological effects predicted in condensed matter physics. Further studies with development of topological models in optics, combined with convenience and flexibility of the photonic platform, allowed to test many models that were proposed but, to the moment, not yet experimentally implemented in condensed matter physics. Examples include the Floquet topological insulators,[7] higher-order topological insulators,[19] and the topological Anderson insulator.[74]

However, a broader potential of topological photonics lays in the fundamental differences between bosonic and fermionic systems enabling to discover and study novel phenomena based on the photonic platform. This includes inherent properties of long-range couplings, the possibility to study non-Hermitian systems with mode leakage and spatially distributed gain and loss, also the possibility to utilize optical nonlinearities for nonlinear effects, thus enriching substantially topological physics with application in photonics. In this context, nonlinear effects in photonic topological systems are of particular interest.[75] In this section, we discuss the recent advances in topological photonics with respect to emission and lasing in small-scale optical structures with a focus on experimental realizations.

Initial developments and achievements in topological lasing were performed in one-dimensional structures of SSH and zigzag arrays. One of the first experimental demonstrations of lasing in topology-protected edge states was implemented in a zigzag lattice of coupled polariton micropillars.[68] Momentum and real-space photoluminescence imaging of the polariton states were performed by non-resonant excitation of the system with a single-mode laser. The measured energy spectrum of the system under uniform excitation represents three bands (S-, P-, D-bands formed by the coupling of corresponding modes of individual micropillar). The coupling of *p*-modes gives rise to two sets of P-bands separated by a finite energy gap. Predominant edge excitation of the system unveils the presence of P-midgap topological edge modes. The measured photoluminescence intensity exhibits nonlinearly increased response indicating topological edge mode lasing regime.

Almost simultaneously with Ref. [68], topological edge-mode lasing was observed in microring resonator arrays.[58, 59] Structures consisted of a 1D linear array of microring resonators (based on InGaAsP quantum well layers) with alternating separations between each other mimicking the SSH model. Weakly coupled termination of the lattice resulted in the emergence of topological edge mode. Introducing gain and loss profile through the mask deposition enables single-topological-mode lasing. However, for higher gain edge mode of the SSH structure starts to compete with bulk ones and exhibits two additional lasing phases.[59]

Single-mode topological edge lasing was also observed in the photonic crystal nanocavity arrays. [60, 61] One of the examples is a linear chain of photonic crystal $L_3$ cavities with SSH-type separations (**Figure 4a**) composed of an InAsP/InP multiple-quantum-well (MQW) epilayer. Lasing excitation spot was chosen to cover the sample partially as shown in the insets of **Figure 4a** by yellow-shaded circles. It allowed to directly control the eigenmode excitation rather bulk or topological edge ones by shifting spot across the structure. The measured photoluminescence spectra for different illumination positions are shown in **Figure 4a**. In the case of the side-system excitation, the peak intensity corresponds to lasing from the topological edge state which was proven by direct visualization of the lasing mode field profile through near-field scanning optical microscopy (**Figure 4a**, bottom).

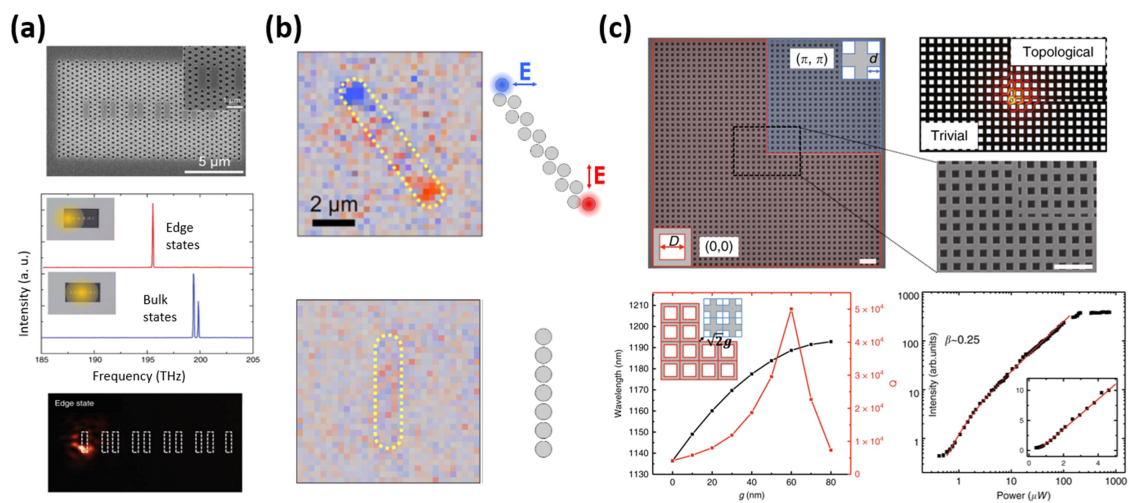

**Figure 4**. **Emission and lasing with topological structures**. (a) Lasing from an array of photonic nanocavities (Reproduced under terms of the CC-BY license.[61] Copyright 2019, Published by Springer Nature); (b) enhanced polarized light emission at Mie-resonant zigzag array of dielectric nanoparticles (Reproduced [78]); (c) lasing at photonic higher-order topological states (Reproduced under terms of the CC-BY license.[83] Copyright 2020, Published by Springer Nature).

Some years ago, there has been a sharp rise of interest in introducing gain materials into topological photonic structures, with the idea to improve semiconductor lasers with the help of topological physics the field of active topological photonics emerged.[76] Examples of remarkable realizations are topological lasers with unidirectional light output under time-reversal symmetry breaking.[76] The cutting-edge research on applications of topological photonics to integrated optical devices can be found in Ref.[77].

Topologically nontrivial nanostructures supported 0D states can also be utilized to study other intriguing effects in light emission. As an example, we mention the paper by Tripathi et al. [78] who demonstrated topologically controlled photoluminescence in zigzag arrays with spin-coated $Er^{3+}$-doped core-shell nanoparticles. Experimentally measured light intensity revealed the enhanced photoluminescence at the topological edge modes. Analysis of the polarization emission manifested strongly dependent polarization emission on the type of topological edge state. **Figure 4b** shows the correspondence between polarizations of topological edge modes of 1D arrays with spatially resolved polarization states of photoluminescence. For topologically trivial chains the effect of enhanced and polarized emission is not observed (see **Figure 4b**, bottom).

Studying topological lasing in photonics is not limited to 1D structures. Thus, rapid developments of two-dimensional photonic topological systems led to the observation of lasing from arbitrary shaped nonreciprocal topological cavities [79] and realization of topological insulator lasers. [80-82] However, in these works, the emission is stimulated by 1D travelling photonic topological edge modes that are out of the scope of our paper. Nevertheless, there is a possibility to construct an active nanophotonic topological cavity hosting 0D localized states. In the paper,[62] the design of such a topological cavity based on dimerized graphene lattice with inversion symmetry breaking was proposed and experimentally investigated. The fabricated sample represented an InGaAsP slab with embedded quantum wells as a gain medium. To explore lasing characteristics the whole structure was optically pumped by a pulsed laser. In the considered valley-Hall system the room-temperature lasing was observed with a narrow spectrum, high coherence, and threshold behavior.[62] Direct observation of spatial field distribution of the light emission manifests the cavity mode lasing above a threshold.

Novel platform of higher-order topological insulators allowed to 'downscale' the eigenmodes and consider 0D topologically protected state within 2D structures. In the work,[83] a topological lasing was observed in 2D photonic crystal nanocavity based on second-order topological corner states at low temperature (4.2K). Such nanocavity can be described by generalized 2D SSH model with considered next-nearest neighbor couplings that break chiral symmetry and open topological band gap. The structure is represented as a photonic crystal GaAs slab with square air holes and composed of two topologically non-equivalent regions where the trivial region wraps around the nontrivial one (**Figure 4c** first row: SEM pictures and electric field profile of the topological corner state). The bulk-edge-corner correspondence guarantees the emergence of second-order topological states at the corners. To observe lasing behavior a single layer of InGaAs quantum dots at the center of the slab was used as the gain. The structure was illuminated at the corner. The distance $g$ between two topologically distinct domains of the structure was optimized to obtain a high Q-factor (equal to 50000 in theory and 5000 in an experiment) while the corner states exhibit a redshift (**Figure 4c** left column, second row). Experimentally retrieved pump-power dependence that shown in **Figure 4c** (right column, second row) demonstrate a low lasing threshold at around 1 μW and a high spontaneous emission coupling factor ($\beta \approx 0.25$). Despite the low operational temperature, the observed threshold is several orders of magnitude lower than of previously realized topological lasers due to the small mode volume and high Q factor, and the performance is comparable to that of conventional nanolasers that, with topological robustness against disorders and defects, provides a great potential of higher-order topology for future nanophotonic devices.

Topology-enhanced emission can be achieved by using halide perovskite nanocrystals. Recently, coupling of halide perovskite nanocrystals to 0D topological corner states was demonstrated for a silicon-based kagome lattice with a layer of perovskite nanocrystals with the emission wavelength precisely tuned to the required wavelength via anion exchange reaction.[84] By measuring the photoluminescence spectra of perovskite nanocrystals, the authors demonstrated a significant emission enhancement at the frequency of zero-dimensional topological corner states.[84]

## 5. Topological Exciton-Polariton Systems

A special type of light-matter quasiparticles known as microcavity polaritons originates from the strong coupling between excitons and photons in a semiconductor microcavity. Polaritons provide an excellent platform for the realisation of many interesting effects including the effects of topological photonics at smaller scales, and in both one- and two-dimensional structures. A typical structure supporting polaritons consists of an optical cavity of a few tens or hundreds of nanometres placed between two Bragg mirrors made of layers of two alternating dielectric materials, as shown in **Figure 3d**. The cavity is designed to be in resonance with the exciton line of embedded semiconductor material. Confined photons excite an exciton as a bound electron-hole pair, which emits a photon that remains trapped in the cavity in the regime of strong light-matter coupling being described as hybrid *cavity polaritons.*[85]

A flexible design of polariton structures allows to study the topological effects for cavity polaritons in one-and two-dimensional potentials.[86] Harder et al.[87] fabricated a zigzag chain of exciton-polariton microcavity traps that, as explained above, can support different types of topological edge states, and they observed highly coherent polariton lasing. They applied an external magnetic field and confirmed directly the excitonic contribution to the polariton robust lasing with high temporal coherence. These experimental results were confirmed by modelling with a generalized Gross−Pitaevskii equation and a Lindblad master equation. More importantly, Harder et al.[87] studied the emission from the topological edge defect by exciting a set of chains of different lengths with an elliptical pumping spot of approximately 30 µm by 3 µm, created by a cylindrical lens. **Figure 5a** shows the structure, and laser-like emission originating from the topological defect of an N = 45 zigzag chain comprising traps with diameters of d = 2 µm and a reduced trap distance of 0.9. The lasing is confirmed by a continuous blue shift of the lasing mode stemming due to the polariton−polariton interaction, see the right image in **Figure 5a**.

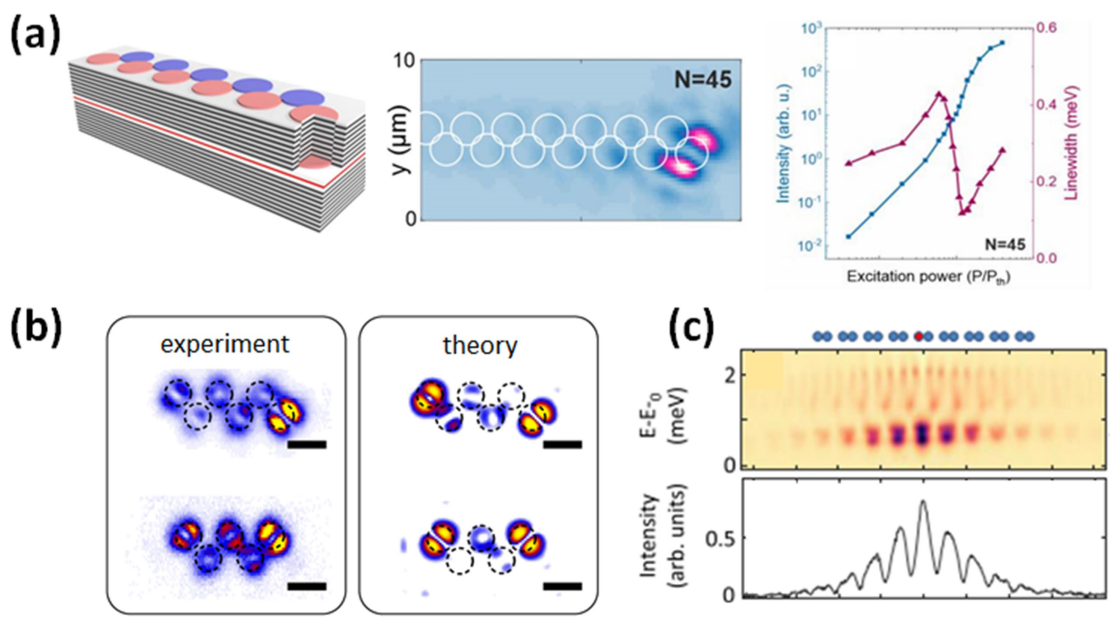

**Figure 5. Topological exciton-polaritons.** (a) Schematic of a zigzag chain created by coupled microcavity traps; observation of the polariton lasing from a topological zigzag chain (N=45), shown also with the integrated intensity and linewidth of the polariton laser emission as a function of the excitation power (Reproduced with permission[87], Copyright 2021, American Chemical Society). (b) Experimental and numerical spatial density distributions of the edge state for the chains of N=6 and N=5 optically induced microcavity traps. The scale bar is 10 µm (Adapted with permission[90], Copyright 2021 The Optical Society). (c) Spatially resolved steady-state emission spectra for the effective SSH lattice and emission intensity as a function of spatial position (Reproduced with permission.[92] Copyright 2021, American Physical Society).

Previous demonstrations of exciton-polariton systems demand cryogenic temperatures and as described above, the study of their topological properties requires the fabrication of many samples. Recently, Su et al.[88] demonstrated a room temperature exciton-polariton topological perovskite zigzag lattice. Like other zigzag lattices, polarization allows to switch between distinct topological phases, and the topologically nontrivial polariton edge states persist in the presence of onsite energy perturbations.

Su et al.[88] demonstrated exciton polariton condensation into the topological edge states under optical pumping. They took advantage of a strong anisotropy and spin-orbit interaction in their samples in the s-band, as predicted earlier by Whittaker et al.[89]. The effective mass (or dispersion) is different for different linear polarisations and, as a result, they induce different tunnelling rates for different polarisations. In contrast, in the studies of Harder et al.[87] and Pieczarka et al.[90], the difference in tunnelling rates is due to the orientation of the p-modes, and the polarisation was not relevant. The spin-orbit interaction lifts the degeneracy between the linearly polarised states.

An all-optical zigzag potential for exciton polaritons has been introduced and realized experimentally by Pieczarka et al.[90] who employed the nonresonant laser excitation and directly measured the topological edge states of a polariton lattice in the regime of bosonic condensation. By using tunability of fully reconfigurable optical lattice, they modified the intersite tunnelling to realize a topological phase transition to a trivial state, in reconfigurable systems that do not require specific sample engineering. This approach allows to study topological phase transitions on demand in fully reconfigurable hybrid systems.

Pieczarka et al.[90] created different chains located at different positions on the sample with similar detuning. The edge state for the $N = 6$ chain is presented in **Figure 5b** with the corresponding results of numerical simulations. For this orientation, the diagonal p-mode configuration is topologically nontrivial. The topological zigzag model with an odd number of sites supports two edge states, however, each edge state comes from a different p-mode configuration. The edge states are also observed for a chain with $N = 5$ sites, as presented in **Figure 5b**. These types of topological localizations resemble the behavior of nonlinear zigzag arrays with competitions of two types of dipole modes excited in each particle.[91] Considering a two-dimensional structure of micropillars, St-Jean et al.[92] combined real- and momentum-space measurements to access directly one-dimensional topological invariants in lattices of semiconductor microcavities confining exciton polaritons. They extracted these invariants in arrays emulating the physics of regular and critically compressed graphene where Dirac cones merge. **Figure 5c** shows spatially resolved steady-state emission spectra for the effective SSH lattice and emission intensity as a function of spatial position. This scheme provides a direct evidence of the bulk-edge correspondence in the polariton systems.

An important generalisation of these results is the study of exciton-polaritons in a zigzag chain of coupled elliptical micropillars subjected to incoherent excitation.[93] This system demonstrates nonreciprocal dynamics and non-Hermitian skin effect, where all the modes of the system collapse to one edge. The polarization splitting and different orientation of the elliptical micropillars allows to predict the topological spin-Hall effect in a one-dimensional lattice, and a phase transition from the Hermitian regime to non-Hermitian regime showing the so-called non-Hermitian skin effect.[94-96]

## 6. Summary and Outlook

It is well established now that topological photonics can offer unique functionalities for manipulation of light at the nanoscale by means of the topological states which are robust against various forms of disorder. Here we have discussed only a few major systems with nontrivial topological phases exhibiting different types of localized states on the nanoscale. Importantly, topological photonics is expected to trigger new fundamental ideas and concepts being at the forefront of research in photonics, as discussed recently.[97]

The study of electronic topological states has a long history. In contrast, topological photonics is a relatively young field of research. Many of the first proof-of-principle demonstrations were reported for microwave frequencies, whereas applications demand the structures operating at the telecom and optical frequencies and having much smaller footprints of topological devices with the target to design and fabricate disorder-immune components for high-speed information processing and optical memory.

As discovered recently, two-dimensional higher-order topological states can display several exotic phenomena such as half-integer charges localized at *disclination defects*[98] with an unconventional higher-order band topology. In addition to the conventional spectral features of gapped edge states and in-gap corner states, a topological band theory predicts that the corner boundary of the higher-order topological insulator can host a 2/3 fractional charge.[99] As the next step, we expect the development of a generalized theory of the mixed geometry-charge response that may provide a uniform approach to the corner and disclination charges. We expect that the predicted effects can be readily observed in experiments and may lead to potential applications in integrated and quantum photonics.

We anticipate that the future development of this field will bring many new discoveries, including deeper understanding of the interplay between *nonlinearity* and *topology*. A nonlinear response can provide a direct way to manipulate topological lattices, and nonlinear response is important for achieving ultrafast modulation. Combining topological photonic structures with nonlinear effects can unlock advanced functionalities such as mode stabilization in lasers, parametric amplifiers protected against feedback, and ultrafast optical switches employing topological waveguides, as discussed recently.[75]

Another important direction is to explore *non-Hermitian systems*, including *topological lasers*. Topological corner states may be useful for selecting and stabilizing lasing modes enabling high-power single-mode operation. Higher-order topological phases enable strong high-quality confinement of light at the corners, and they support lasing action for several multipole corner modes with distinct emission profiles revealed via hyperspectral imaging and discern signatures of non-Hermitian radiative coupling of leaky topological states.[100] Such modes can demonstrate selective lasing from edge and corner states within the topological bandgap.

In addition, non-Hermitian topological systems can be modified by *temporal modulation*. As was discovered recently, the nontrivial topological features in the energy bands of non-Hermitian systems provide promising pathways to achieve robust physical behaviors in open systems. Wang et al.[101] demonstrated experimentally nontrivial winding by implementing non-Hermitian lattice Hamiltonians along a frequency synthetic dimension formed in a ring resonator undergoing simultaneous phase and amplitude modulations. The authors demonstrated that the topological winding can be controlled by changing the modulation waveform. Their results allow for the synthesis and characterization of topologically nontrivial phases in nonconservative systems.

One of the important expansions of the physics of topological phases can be found in *quantum systems* which we did not discuss in this paper, see e.g., Refs. [102-104] Here we mention only a few recent demonstrations relevant to the models introduced above. Klauck et al. [103] studied two-particle quantum correlations of photons in the SSH array of optical waveguides, and they found that, whereas at the trivial edge in the SSH lattice the bunching behavior of the indistinguishable photons remains unchanged, the bunching is distorted at the topological edge. Thus, the existence of the topological edge in the SSH model may have a significant effect on the quantum properties and interference of indistinguishable photons. Several quantum phenomena appearing in a topological waveguide QED system described by the SSH model have been discussed by Bello et al.[104] who predicted the appearance of chiral photon bound states with topological robustness mediating directional, long-range spin interactions, leading to exotic phases.

Finally, it is interesting to mention that many effects associated with topological phases revealed for optical systems and described by either SSH or zigzag arrays can occur in completely different systems such as the antisymmetric Lotka-Volterra equation describing the evolutionary dynamics of *a rock-paper-scissors cycle*. A chain of rock-paper-scissor cycles resembles our zigzag array of nanoparticles, with topological phases manifested as robust polarization states.[105]

**Acknowledgements**

Y.K. acknowledges useful discussions of topological exciton-polaritons with Alberto Amo, Eliezer Estrecho, Maciej Pieczarka, and Qihua Xiong, and he also thanks Sergey Kruk, Alexey Slobozhanyuk, and Alexander Poddubny for useful collaboration on the topics of topological photonics. This work was supported by the Australian Research Council (grant DP200101168), and the Russian Science Foundation (grant no. 21-72-30018).

**References**

[1] M.I. Stockman, *Opt. Express* **2011**, *19*, 22029;

[2] K. Koshelev, Y. Kivshar, *ACS Photonics* **2021**, *8*, 102;

[3] F. D. M. Haldane, S. Raghu, *Phys. Rev. Lett.* **2008**, *100*, 013904;

[4] Z. Wang, Y. Chong, J.D. Joannopoulos, M. Soljacic, *Nature* **2009**, *461*, 772;

[5] M. Hafezi, E.A. Demler, M.D. Lukin, J.M. Taylor, *Nat. Phys.* **2011**, *7*, 907;

[6] A.B. Khanikaev, S.H. Mousavi, W.-K. Tse, M. Kargarian, A.H. MacDonald, G. Shvets, *Nat. Mater.* **2013**, *12*, 233;

[7] M.C. Rechtsman, J.M. Zeuner, Y. Plotnik, Y. Lumer, D. Podolsky, F. Dreisow, S. Nolte, M. Segev, A. Szameit, *Nature* **2013**, *496*, 196;

[8] M. Hafezi, S. Mittal, J. Fan, A. Migdall, J.M. Taylor, Nat. Photonics **2013**, 7, 1001–1005;

[9] L. Lu, J.D. Joannopoulos, M. Soljačić, *Nat. Photonics* **2014**, *8*, 821;

[10] X. Cheng, C. Jouvaud, X. Ni, S.H. Mousavi, A.Z. Genack, A.B. Khanikaev, *Nat. Mater.* **2016**, *15*, 542;

[11] T. Ozawa, H.M. Price, A. Amo, N. Goldman, M. Hafezi, L. Lu, M.C. Rechtsman, D. Schuster, J. Simon, O. Zilberberg, I. Carusotto, *Rev. Mod. Phys.* **2019**, *91*, 015006;

[12] W.P. Su, J.R. Schrieffer, A.A. Heeger, *Phys. Rev. Lett.* **1979**, *42*, 1698;

[13] A.J. Heeger, S. Kivelson, J.R. Schrieffer, W.P. Su, *Rev. Mod. Phys.* **1988**, *60*, 781;

[14] H.T. Schomerus, *Opt. Lett.* **2013**, *38*, 1912;

[15] S.R. Pocock, X. Xiao, P.S. Huidobro, V. Giannini, Topological Plasmonic Chain with Retardation and Radiative Effects, *ACS Photonics* **2018**, *5*, 2271; https://pubs.acs.org/doi/full/10.1021/acsphotonics.8b00117;

[16] A. Poddubny, A. Miroshnichenko, A. Slobozhanyuk, Y. Kivshar, *ACS Photonics* **2014**, *1*, 101;

[17] I. Sinev, I. Mukhin, A. Slobozhanyuk, A. Poddubny, A. Miroshnichenko, A. Samusev, Y.S. Kivshar, *Nanoscale* **2015**, *7*, 11904;

[18] A.P. Slobozhanyuk, A.N. Poddubny, A.E. Miroshnichenko, P.A. Belov, Y.S. Kivshar, *Phys. Rev. Lett.* **2015**, *114*, 123901;

[19] W.A. Benalcazar, B.A. Bernevig, T.L. Hughes, *Science* **2017**, *357*, 61;

[20] W.A. Benalcazar, B.A. Bernevig, T.L. Hughes, *Phys. Rev. B* **2017**, *96*, 245115;

[21] Z. Song, Z. Fang, C. Fang, *Phys. Rev. Lett.* **2017**, *119*, 246402;

[22] F. Schindler, A.M. Cook, M.G. Vergniory, Z. Wang, S.P. Parkin, A. Bernevig, T. Neupert, *Sci. Adv.* **2018**, *4*, eaat0346;

[23] M. Ezawa, *Phys. Rev. Lett.* **2018**, *120*, 026801;

[24] M. Serra-Garcia, V. Peri, R. Susstrunk, O.R. Bilal, T. Larsen, L.G. Villanueva, S.D. Huber, *Nature* **2018**, *555*, 342;

[25] X. Ni, M. Weiner, A. Alù, A.B. Khanikaev, *Nat. Mater.* **2019**, *18*, 113;

[26] S. Imhof, C. Berger, F. Bayer, J. Brehm, L.W. Molenkamp, T. Kiessling, F. Schindler, C.H. Lee, M. Greiter, T. Neupert, R. Thomale, *Nat. Phys.* **2018**, *14*, 925;

[27] C.W. Peterson, W.A. Benalcazar, T.L. Hughes, G. Bahl, *Nature* **2018**, *555*, 346;

[28] J. K. Asboth, L. Oroszlany, A. Palyi, *A Short Course on Topological Insulators: Band-structure topology and edge states in one and two dimensions*, Springer, Lecture Notes in Physics, 919 **2016**.

[29] J. Zak, Phys. Rev. Lett. **1989**, 62, 2747;

[30] R. Jackiw, C. Rebbi, *Phys. Rev. D* **1976**, *13*, 3398;

[31] L. Li, Z. Xu, S. Chen, *Phys. Rev. B* **2014**, *89*, 085111;

[32] Y. Hadad, A.B. Khanikaev, A. Alù, *Phys. Rev. B* **2016**, *93*, 155112;

[33] Y. Hadad, J.C. Soric, A.B. Khanikaev, A. Alù, *Nat. Electron.* **2018**, *1*, 178;

[34] D.A. Dobrykh, A.V. Yulin, A.P. Slobozhanyuk, A.N. Poddubny, Y.S. Kivshar, *Phys. Rev. Lett.* **2018**, *121*, 163901;

[35] M. Di Liberto, A. Recati, I. Carusotto, C. Menotti, *Phys. Rev. A* **2016**, *94*, 062704;

[36] M.A. Gorlach, A.N. Poddubny, *Phys. Rev. A* **2017**, *95*, 053866;

[37] A.P. Slobozhanyuk, A.N. Poddubny, I.S. Sinev, A.K. Samusev, Y.F. Yu, A.I. Kuznetsov, A.E. Miroshnichenko, Y.S. Kivshar, *Laser Photonics Rev.* **2016**, *10*, 656;

[38] S. Kruk, A. Slobozhanyuk, D. Denkova, A. Poddubny, I. Kravchenko, A. Miroshnichenko, D. Neshev, Y. Kivshar, *Small* **2017**, *13*, 1603190;

[39] L. Lin, S. Kruk, Y. Ke, C. Lee, Y. Kivshar, Topological states in disordered arrays of dielectric nanoparticles, *Phys. Rev. Res.* **2020**, *2*, 043233; https://doi.org/10.1103/PhysRevResearch.2.043233;

[40] S. Mittal, V.V. Orre, G. Zhu, M.A. Gorlach, A. Poddubny, M. Hafesi, *Nat. Photonics* **2019**, *13*, 692;

[41] A. El Hassan, F.K. Kunst, A. Moritz, G. Andler, E.J. Bergholtz, M. Bourennane, *Nat. Photonics* **2019**, *13*, 697;

[42] M. Li, D. Zhirihin, M. Gorlach, X. Ni, D. Filonov, A. Slobozhanyuk, A. Alù, A.B. Khanikaev, *Nat. Photonics* **2020**, *14*, 89;

[43] S. Kruk, W. Gao, D.-Y. Choi, T. Zentgraf, S. Zhang, Y. Kivshar, *Nano Lett.* **2021**, *21*, 4592;

[44] W.A. Benalcazar, T. Li, T.L. Hughes, *Phys. Rev. B* **2019**, *99*, 245151;

[45] X. Ni, M.A. Gorlach, A. Alu, A.B. Khanikaev, *New J. Phys.* **2017**, *19*, 055002;

[46] A. Vakulenko, S. Kiriushechkina, M. Wang, M. Li, D. Zhirihin, X. Ni, S. Guddala, D. Korobkin, A. Alu, A.B. Khanikaev, *Adv. Mater.* **2021**, *33*, 2004376;

[47] C. L. Kane, T. C. Lubensky, *Nat. Phys.* **2014**, *10*, 39;

[48] R. Susstrunk, S.D. Huber, *Science* **2015**, *349*, 47;

[49] S.H. Mousavi, A.B. Khanikaev, Z. Wang, *Nat. Commun.* **2015**, *6*, 8682;

[50] S.D. Huber, Topological mechanics. *Nat. Phys.* **2016**, *12*, 621;

[51] Z. Yang, F. Gao, X. Shi, X. Lin, Z. Gao, Y. Chong, B. Zhang, *Phys. Rev. Lett.* **2015**, *114*, 114301;

[52] A.B. Khanikaev, R. Fleury, S.H. Mousavi, A. Alù, *Nat. Commun.* **2015**, *6*, 8260;

[53] R. Fleury, A. Khanikaev, A. Alu, *Nat. Commun.* **2016**, *7*, 11744;

[54] Y.-G. Peng, C.-Z. Qin, D.-G. Zhao, Y.-X. Shen, X.-Y. Xu, M. Bao, H. Jia, X.-F. Zhu, *Nat. Commun.* **2016**, *7*, 13368;

[55] C. W. Ling, Meng Xiao, C. T. Chan, S. F. Yu, K. H. Fung, *Opt. Express* **2015**, *23*, 2021;

[56] M. Proctor, M. B. de Paz, D. Bercioux, A. García-Etxarri, and P.A. Huidobro. *Appl. Phys. Lett.* **2021**, *118*, 091105;

[57] Y. Chen, Z.K. Lin, H. Chen, J.H. Jiang, *Phys. Rev. B* **2020**, *101*, 041109;

[58] H. Zhao, P. Miao, M.H. Teimourpour, S. Malzard, R. El-Ganainy, H. Schomerus, L. Feng, Topological hybrid silicon microlasers. *Nat. Commun.* **2018**, *9*, 981; https://doi.org/10.1038/s41467-018-03434-2;

[59] M. Parto, S. Wittek, H. Hodaei, G. Harari, M.A. Bandres, J. Ren, M.C. Rechtsman, M. Segev, D.N. Christodoulides, M. Khajavikhan, *Phys. Rev. Lett.* **2018**, *120*, 113901;

[60] Y. Ota, R. Katsumi, K. Watanabe, S. Iwamoto, Y. Arakawa, Topological photonic crystal nanocavity laser. *Commun. Phys.* **2018**, 1, 86; https://doi.org/10.1038/s42005-018-0083-7;

[61] C. Han, M. Lee, S. Callard, C. Seassal, H. Jeon, Lasing at topological edge states in a photonic crystal L3 nanocavity dimer array. *Light: Sci. Appl.* **2019**, *8*, 40; https://doi.org/10.1038/s41377-019-0149-7;

[62] D. Smirnova, A. Tripathi, S. Kruk, M.S. Hwang, H.R. Kim, H.G. Park, Y. Kivshar, *Light: Sci. Appl.* **2020**, *9*, 127;

[63] Y. Ota, F. Liu, R. Katsumi, K. Watanabe, K. Wakabayashi, Y. Arakawa, S. Iwamoto, *Optica* **2019**, *6*, 786;

[64] D. Smirnova, S. Kruk, D. Leykam, E. Melik-Gaykazyan, D.Y. Choi, Y. Kivshar, *Phys. Rev. Lett.* **2019**, *123*, 103901;

[65] T. Karzig, C.-E. Bardyn, N.H. Lindner, G. Refael, *Phys. Rev. X* **2015**, *5*, 031001;

[66] A. V. Nalitov, D. D. Solnyshkov, G. Malpuech, *Phys. Rev. Lett.* **2015**, *114*, 116401;

[67] D.V. Solnyshkov, A.V. Nalitov, G. Malpuech, *Phys. Rev. Lett.* **2016**, *16*, 046402;

[68] P. St-Jean, V. Goblot, E. Galopin, A. Lemaître, T. Ozawa, L. Le Gratiet, I. Sagnes, J. Bloch, A. Amo, *Nat. Photonics* **2017**, *11*, 651;

[69] S. Klembt, T. H. Harder, O. A. Egorov, K. Winkler, R. Ge, M. A. Bandres, M. Emmerling, L. Worschech, T. C. H. Liew, M. Segev, C. Schneider, S. Höfling, *Nature* **2018**, *562*, 552;

[70] Y.V. Kartashov, D.V. Skryabin, *Phys. Rev. Lett.* **2019**, *122*, 083902;

[71] W. Liu, Z. Ji, Y. Wang, G. Modi, M Hwang, B. Zheng, V.J. Sorger, A. Pan, R. Agarwal, *Science* **2020**, *370*, 6516, 600;

[72] M. Li, I. Sinev, F. Benimetskiy, T. Ivanova, E. Khestanova, S. Kiriushechkina, A. Vakulenko, S. Guddala, M. Skolnick, V.M. Menon, D. Krizhanovskii, A. Alù, A. Samusev, A.B. Khanikaev, *Nat. Commun.* **2021**, *12*, 4425;

[73] D. D. Solnyshkov, G. Malpuech, P. St-Jean, S. Ravets, J. Bloch, A. Amo, *Optics Materials Express* **2021**, *11*, 1119;

[74] S. Stützer, Y. Plotnik, Y. Lumer, P. Titum, N.H. Lindner, M. Segev, M.C. Rechtsman, A. Szameit, *Nature* **2018**, *560*, 461;

[75] D. Smirnova, D. Leykam, Y. Chong, Y. Kivshar, *Appl. Phys. Rev.* **2020**, *7*, 021306;

[76] Y. Ota, K. Takata, T. Ozawa, A. Amo, Z. Jia, B. Kante, M. Notomi, Y. Arakawa, S. Iwamoto, *Nanophotonics* **2020**, *9*, 547;

[77] Y. Wu, C. Li, X. Hu, Y. Ao, Y. Zhao, Q. Gong, *Adv. Opt. Mater.* **2017**, *5*, 1700357;

[78] A. Tripathi, S. Kruk, Y. Shang, J. Zhou, I. Kravchenko, D. Jin, Y. Kivshar, *Nanophotonics* **2020**, *10*, 435;

[79] B. Bahari, A. Ndao, F. Vallini, A. El Amili, Y. Fainman, B. Kanté, *Science* **2017**, *358*, 636;

[80] M.A. Bandres, S. Wittek, G. Harari, M. Parto, J. Ren, M. Segev, D.N. Christodoulides, M. Khajavikhan, *Science* **2018**, *359*, 6381, eaar4005;

[81] G. Harari, M. A. Bandres, Y. Lumer, M.C. Rechtsman, Y. D. Chong, M. Khajavikhan, D.N. Christodoulides, M. Segev, *Science* **2018**, *359*, 6381, eaar4003;

[82] Y. Zeng, U. Chattopadhyay, B. Zhu, B. Qiang, J. Li, Y. Jin, L. Li, A.G. Davies, E.H. Linfield, B. Zhang, Y. Chong, Q. Jie, Wang, *Nature* **2020**, *578*, 246;

[83] W. Zhang, X. Xie, H. Hao, J. Dang, S. Xiao, S. Shi, H. Ni, Z. Niu, C, Wang, K. Jin, X. Zhang, X. Xu, Low-threshold topological nanolasers based on the second-order corner state. *Light: Sci. Appl.* **2020**, *9*, 109; https://doi.org/10.1038/s41377-020-00352-1;

[84] A.S. Berestennikov, A. Vakulenko, S. Kiriushechkina, M. Li, Y. Li, L.E. Zelenkov, A.P. Pushkarev, M.A. Gorlach, A.L. Rogach, S.V. Makarov, A.B. Khanikaev, *J. Phys. Chem. C* **2021**, *125*, 9884;

[85] A. Kavokin, J.J. Baumberg, G. Malpuech, F.P. Laussy, *Microcavities,* Oxford University Press Inc., New York **2007**.

[86] S. Klembt, T.H. Harder, O.A. Egorov, K. Winkler, R. Ge, M.A. Bandres, M. Emmerling, L. Worschech, T.C.H. Liew, M. Segev, C. Schneider, S. Hofling, *Nature* **2018**, *562*, 552;

[87] T.H. Harder, M. Sun, O.A. Egorov, I. Vakulchyk, J. Beierlein, P. Gagel, M. Emmerling, C. Schneider, U. Peschel, I.G. Savenko, S. Klembt, S. Höfling, *ACS Photonics* **2021**, *8*, 1377;

[88] R. Su, S. Ghosh, T.C.H. Liew, Q. Xiong, *Sci. Adv.* **2021**, *7*, eabf8049;

[89] C.E. Whittaker, E. Cancellieri, P.M. Walker, B. Royall, L.E. Tapia Rodriguez, E. Clarke, D. M. Whittaker, H. Schomerus, M.S. Skolnick, D.N. Krizhanovskii, *Phys. Rev B* **2019**, *99*, 081402(R);

[90] M. Pieczarka, E. Estrecho, S. Ghosh, M. Wurdack, M. Steger, D.W. Snoke, K. West, L.N. Pfeiffer, T.C.H. Liew, A.G. Truscott, E.A. Ostrovskaya, Topological phase transition in an all-optical exciton-polariton lattice, *Optica* **2021**, *8*, 1084; https://doi.org/10.1364/OPTICA.426996;

[91] S. Kruk, A. Poddubny, D. Smirnova, L. Wang, A. Slobozhanyuk, A. Shorokhov, I. Kravchenko, B. Luther-Davies, Y. Kivshar, *Nat. Nanotechnol.* **2019**, *14*, 126;

[92] P. St-Jean, A. Dauphin, P. Massignan, B. Real, O. Jamadi, M. Milicevic, A. Lemaître, A. Harouri, L. Le Gratiet, I. Sagnes, S. Ravets, J. Bloch, A. Amo, *Phys. Rev. Lett.* **2021**, *126*, 127403;

[93] S. Mandal, R. Banerjee, E. A. Ostrovskaya, T.C.H. Liew, *arXiv:2009.00924v2* **2020**.

[94] S. Mandal, R. Banerjee, T.C.H. Liew, *arXiv:2103.05480v1* **2021**.

[95] N. Hatano, D.R. Nelson, *Phys. Rev. Lett.* **1996**, *77*, 570;

[96] S. Yao, Z. Wang, *Phys. Rev. Lett.* **2018**, *121*, 086803;

[97] M. Segev, M.A. Bandres, *Nanophotonics* **2021**, *10*, 425;

[98] Y. Liu, S. Leung, F.-F. Li, Z.-K. Lin, X. Tao, Y. Poo, J.-H. Jiang, *Nature* **2021**, *589*, 381;

[99] S. Wu, B. Jiang, Y. Lai, J.-H. Jiang, *Photonics Res.* **2021**, *9*, 668-677;

[100] H.-R. Kim, M.-S. Hwang, D. Smirnova, K.-Y. Jeong, Y. Kivshar, H.-G. Park, *Nat. Commun.* **2020**, *11*, 5758;

[101] K. Wang, A. Dutt, K.Y. Yang, C.C. Wojcik, J. Vuckovic, S. Fan, *Science* **2021**, *371*, 1240;

[102] A. Blanco-Redondo, B. Bell, D. Oren, B.J. Eggleton, M. Segev, *Science* **2018**, *362*, 568;

[103] F. Klauck, M. Heinrich, A. Szameit, *Photonics Res.* 2021, *9,* A1;

[104] M. Bello, G. Platero, J. I. Cirac, A. González-Tudela, *Sci. Adv.* 2019, *5*, eaaw0297;

[105] J. Knebel, P.M. Geiger, E. Frey, *Phys. Rev. Lett.* 2020, *125, 258301.*

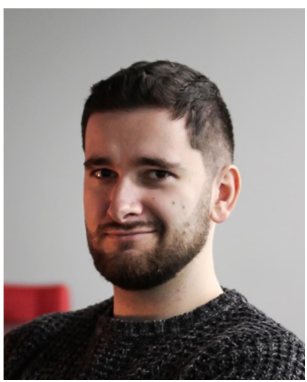

**Dmitry V. Zhirihin** was born in St. Petersburg, Russia, in 1995. He received BSc in 2015 from Peter The Great St. Petersburg Polytechnic University, M.Sc. degree in photonics (cum laude) in 2017 from the ITMO University, and PhD in electromagnetic engineering from the ITMO University in 2020. Currently, Dmitry is Junior Research Fellow at Faculty of Physics, Department of Physics and Engineering of the ITMO University. His main research interests incude topological photonics, applied electromagnetics, metamaterials, metasurfaces, antennas, and microwave experiments.

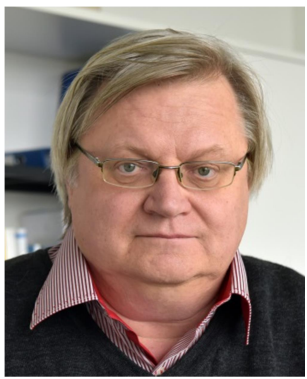

**Yuri S. Kivshar** is Distinguished Professor of the Australian National University working in photonics and metamaterials. He received PhD degree in 1984 in Kharkov (Ukraine), and in 1989 he left the Soviet Union to work in Spain and then later in Germany, as a Humboldt Research Fellow. In 1993, Yuri Kivshar moved to Australia where he established Nonlinear Physics Center. He is one of the founders of all-dielectric resonant metaphotonics (or "Mie-tronics"), which is governed by Mie resonances in high-index dielectric nanoparticles. Yuri Kivshar is Fellow of the Australian Academy of Science, OSA, APS, SPIE, and IOP.

**D.V. Zhirihin, Y.S. Kivshar**

**Topological Photonics on a Small Scale**

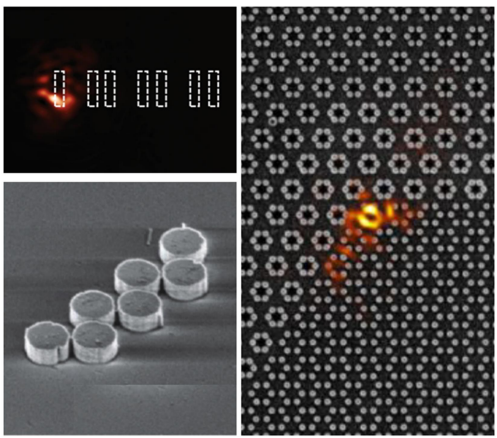 In this Perspective, Dmitry Zhirihin and Yuri Kivshar discuss the recent advances in small-scale topological photonics based on low-dimensional localized states supported by subwavelength optical structures. They summarize the key achievements in this field and provide an outloook of the future progress of small-scale topological nanophotonics.